# Effects of high energy proton irradiation on the superconducting properties of Fe(Se,Te) thin films

G. Sylva[1], E. Bellingeri[1], C. Ferdeghini[1], A. Martinelli[1], I. Pallecchi[1], L. Pellegrino[1], M. Putti[1,2], G. Ghigo[3], L. Gozzelino[3], D. Torsello[3], G. Grimaldi[4], A. Leo[4,5], A. Nigro[4,5], V. Braccini[1,*]

[1] National Research Council CNR-SPIN Genova, C.so Perrone 24, I-16152 Genova, Italy

[2] Physics Department, University of Genova, via Dodecaneso 33, 16146 Genova, Italy

[3] Department of Applied Science and Technology, Politecnico di Torino and INFN Sezione di Torino, C.so Duca degli Abruzzi 24, I-10129 Torino, Italy.

[4] National Research Council CNR-SPIN Salerno, Via Giovanni Paolo II 132, I-84084 Fisciano (SA), Italy

[5] Physics Department 'E.R. Caianiello', University of Salerno, Via Giovanni Paolo II 132, I-84084 Fisciano (SA), Italy

* valeria.braccini@spin.cnr.it

**Abstract**

In this paper we explore the effects of 3.5 MeV proton irradiation on Fe(Se,Te) thin films grown on $CaF_2$. In particular, we carry out a systematic experimental investigation with different irradiation fluences up to $7.30 \cdot 10^{16}$ cm$^{-2}$ and different proton implantation depths, in order to clarify whether and to what extent the critical current is enhanced or suppressed, what are the effects of irradiation on the critical temperature, the resistivity and the critical magnetic fields, and finally what is the role played by the substrate in this context.

We find that the effect of irradiation on superconducting properties is generally small as compared to the case of other iron-based superconductors. Such effect is more evident on the critical current density $J_c$, while it is minor on the transition temperature $T_c$, on the normal state resistivity $\rho$ and on the upper critical field $H_{c2}$ up to the highest fluences explored in this work. In addition, our analysis shows that when protons implant in the substrate far from the superconducting film, the critical current can be enhanced up to 50% of the pristine value at 7 T and 12 K, while there is no appreciable effect on critical temperature and critical fields together with a slight decrease in resistivity. On the contrary, when the implantation layer is closer to the film-substrate interface, both critical current and temperature show a decrease accompanied by an enhancement of the resistivity and the lattice strain. This result evidences that possible modifications induced by irradiation in the substrate may affect the superconducting properties of the film via lattice strain. The robustness of the Fe(Se,Te) system to irradiation induced damage makes it a promising compound for the fabrication of magnets in high-energy accelerators.

**Introduction**

Effects of irradiation on iron-based superconductors (FeSCs) have been investigated since earlier stages after their discovery with several goals, namely gaining information on fundamental properties such as gap symmetry and suppression of the critical temperature $T_c$ by impurity scattering [1,2], investigating vortex physics and flux pinning in view of applications in magnet fabrication [3], and testing the robustness or deterioration of superconducting properties by irradiations for application in high-energy accelerators.

Given the unconventional pairing mechanism and the symmetry of the order parameter in FeSCs, it was initially expected that irradiation damage would suppress the superconducting properties significantly. On the contrary, only mild $T_c$ suppression was observed and visible enhancement of the critical current density $J_c$ were observed in the so-called 122 and 1111 FeSCs families (see ref. [3] for a review). In particular, in the 1111 family, a $J_c$ values up to $2 \cdot 10^7$ A/cm$^2$ were observed after heavy-ion irradiation [4].

In the 11 family, different results have been reported depending on the phase, the type of particle and their energy. For example, thin films grown on $LaAlO_3$ by our group and irradiated with neutrons, duplicated their critical current at 15 K with no change in $T_c$ [5]. Conversely, an increase of $T_c$ was found after neutron [6], electron [7] and proton [8] irradiation. This remarkable result highlights a mechanism for an increase in transition temperature which can overcompensate the detrimental effect of disorder on $T_c$. Surprising results have been found specifically in FeSe [7] and Fe(Se,Te) [8] samples. In ref. [7], a $T_c$ enhancement of 0.4 K upon electron irradiation was detected in FeSe single crystals, and was interpreted by the authors as due to local strengthening of the magnetic pairing mechanism by irradiation-induced defects. In Fe(Se,Te) thin films irradiated by low-energy protons, not only $J_c$ (T=4.2K, H=0) was increased by 55% and pinning force was increased in the high-field regime upon irradiation, but remarkably a simultaneous $T_c$ enhancement by 0.5 K was detected [8]. The authors of ref. [8] consider possible mechanisms for $T_c$ enhancement in Fe(Se,Te) films, namely phonon-related interface effects [9], chemical effects related to excess Fe at interstitial sites and Se/Te ratio [10] and strain effects [11], and on the basis of structural analysis they interpret their results in terms of coexistence of nanoscale regions subject to compressive and tensile strain, originated by the irradiation defects.

On the other hand, a more extended investigation carried out on heavy-ion irradiated FeSe$_{0.4}$Te$_{0.6}$ single crystals yielded to not systematic and not reproducible results on superconducting properties [12], even if a significant effect in terms of vortex-pinning enhancement on FeSe$_{0.45}$Te$_{0.55}$ crystals has been proven on the atomic scale [13]. Recently, on the contrary, Fe(Se,Te) thin films grown on CaF$_2$ were reported to show a decrease in T$_c$ up to 7 K upon irradiation with 3.5 MeV protons [14].

Clearly, in advance of conjecturing on the mechanisms for T$_c$ and J$_c$ enhancements, understanding irradiation effects in 11 FeSCs requires a larger systematic experimental investigation carried out with different fluences and energies, as well as a comprehensive comparison of samples available in literature, correlating the experimental results with the expected disorder created in the sample by irradiation.

In this paper we aim to address several questions concerning the effects of 3.5 MeV proton irradiation of FeSe$_{0.5}$Te$_{0.5}$ films, namely whether and to what extent the critical current can be enhanced, what is the effect on the strain, on the critical temperature and on the critical magnetic fields, and finally what is the role played by the substrate. The latter is a crucial and complex issue, unavoidable in the case of thin films: besides defects created in the film, impinging protons produce modifications of the substrate that in turn have influence on the properties of the film. To address this issue, we compare results obtained in different experimental conditions, i.e. same defect density created by protons in the film and different proton-implantation depths into the substrate, obtained by lowering the proton energy with the interposition of a thin Aluminium foil between the proton beam and the sample.

**Experimental details**

The thin films used for this irradiation experiment were deposited on oriented 001 CaF$_2$ single crystals in an ultra-high vacuum PLD system equipped with a Nd:YAG laser at 1024 nm using a FeSe$_{0.5}$Te$_{0.5}$ target prepared by direct synthesised with a two-step method [15]. The deposition was carried out at a residual gas pressure of 10$^{-8}$ mbar while the substrate was kept at 350 °C. The parameters of the laser used during the deposition are 3 Hz as laser repetition rate, 2 J/cm$^2$ as laser fluency (2 mm$^2$ spot size) and 5 cm as distance between target and sample, and have been optimised to obtain high quality epitaxial Fe(Se,Te) thin films [16].

Five films 100 nm thick were prepared for irradiation. Three of them, indicated in the following as samples A, B and C, were patterned and designated for transport measurements while samples D and E were analysed with X-ray diffraction. Before irradiation, all the films were analysed with X-ray diffraction using a four-circle diffractometer. This analysis confirms the phase purity of all films and the optimum epitaxial growth of all the films. Φ scans reveal that films grow rotated by 45° with respect to the *a* axis due to the good matching with the half of the diagonal of CaF$_2$ crystalline cell, as already reported [17].

By using the *00l* diffraction patterns, the values of the *c* cell parameter for the different samples were evaluated by means of Rietveld refinement with the program Fullprof. At this scope, the instrumental resolution function and the zero-shift parameters were refined using the CaF$_2$ substrate as reference; diffraction lines were modelled by a Thompson-Cox-Hastings pseudo-Voigt convoluted with axial divergence asymmetry function. The so obtained values were fixed and, imposing a full *c*-axis texturing, the parameters pertaining to the FeSe$_{0.5}$Te$_{0.5}$ thin films were refined, i.e. the lattice parameter *c* and the Lorentzian strain parameter. Lattice micro-strain along (*00l*) was evaluated by the refined strain parameters and analysing the broadening of diffraction lines by means of the Williamson-Hall plot method [18]. Generally, in the case where size effects are negligible and the micro-strain is isotropic, a straight line passing through all the points in the plot and through the origin has to be observed, where the slope provides the micro-strain: the higher the slope the higher the micro-strain. If the broadening is not isotropic, size and strain effects along particular crystallographic directions can be obtained by considering different orders of the same reflection. In the present case, all the analysed diffraction lines pertain to the same *00l* family of planes.

In order to allow the measurement of critical current, the films designated for transport measurements were patterned through standard optical photolithography and the etching performed by water-cooled Argon ion milling (Argon ions energy 500 eV). After the milling process, the photoresist was removed by mild sonication in acetone for few tens of seconds and dried in Nitrogen air blow. Nine Hall bar shaped micro bridges of 20x50 µm² size were realized (see Figure 1). Single Hall bars or groups of Hall bars were then selectively irradiated with different fluence values as detailed in the following.

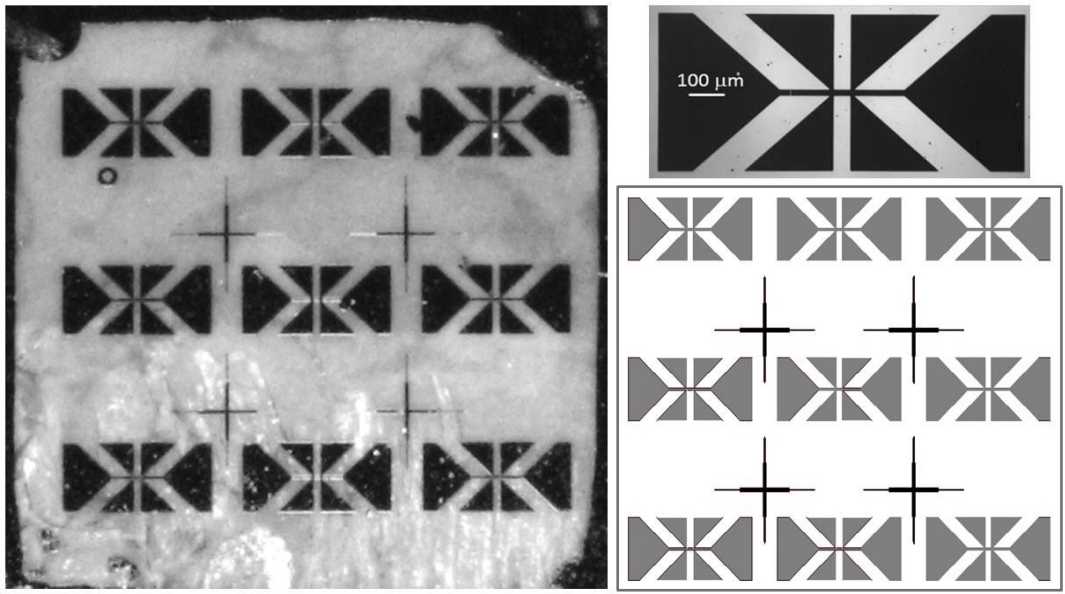

**Figure 1**: Optical viewgraph of a Fe(Se,Te) thin film sample showing the nine Hall bar shaped micro bridges (left) and scheme of the mask pattern(right).

The films were irradiated with 3.5 MeV protons at the CN Van de Graaf accelerator of INFN-LNL (Istituto Nazionale di Fisica Nucleare - Laboratori Nazionali di Legnaro, Italy). The ion beam was parallel to the *c*-axis of the films and the proton flux was always kept below to $10^{12}$ cm$^{-2}$s$^{-1}$ in order to minimize the heating of the samples under irradiation.

In order to investigate the influence of the proton's implantation depth on the structural and electrical properties of the films, some of the samples were directly irradiated with 3.5 MeV protons while others were irradiated with protons decelerated through the interposition of an 80-µm thick Aluminium foil. Details of the irradiation experiment (e.g. the choice of the Al-foil thickness, the adopted fluences) were guided by previous accurate simulations of the damage induced both in the film and in the substrate, obtained by the Monte Carlo codes SRIM [19]. In all cases, protons crossed the films and implanted into the substrate. The implantation depths in the CaF$_2$ substrate are 86 µm without Al foil and 21 µm with the 80 µm thick Al foil. The implantation profiles are shown in Figure 2. Protons are expected to produce random point defects and some defect nanoclusters in the film/substrate [20], due to the Coulomb scattering with atomic nuclei. SRIM calculations predict a homogeneous energy release along the superconducting films thickness (100 nm). The Bragg peak is located in the substrate, in correspondence of the implantation peak shown in Figure 2. Following ref. [21], we estimated the expected damage in the film, in terms of displacements per atom (d.p.a.), using the modified

Kinchin Pease approach [22]. The average values of d.p.a. in the $FeSe_{0.5}Te_{0.5}$ films are summarized in Table 1 for each irradiation experiment together with the lower limit of the average distance between proton-induced defects inferred by d.p.a., disregarding possible Frenkel pairs clustering and annealing effects.

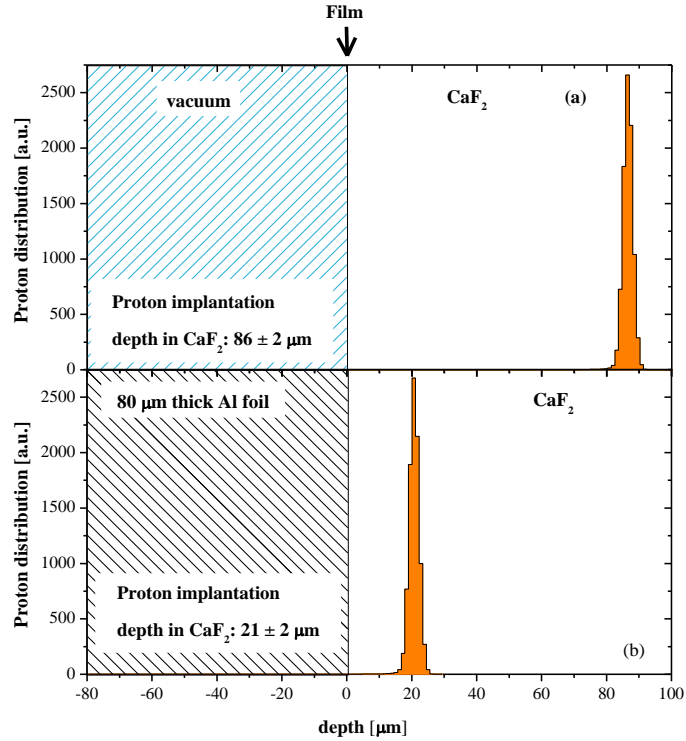

**Figure 2**: Spatial distribution of the implanted protons in the $CaF_2$ substrate without (a) and with (b) the interposition of an Al foil to decelerate the protons. The zero depth corresponds to the sample surface. The sample consists of a 100 nm thick $FeSe_{0.5}Te_{0.5}$ (not visible) and its $CaF_2$ substrates.

Electrical transport properties of the micro bridges as a function of temperature and magnetic field were measured in a Physical Properties Measuring System (PPMS) by Quantum Design up to 9 T and in a Cryogenic-Free Measurement System (CFM) by Cryogenic Ltd. up to 16 T. Resistivity measurements as function of the applied magnetic field have been performed by standard four probe current-biased measurement technique. Critical current values at different temperatures and magnetic fields were extracted from voltage versus current characteristics acquired in the PPMS system by sweeping the current from zero with exponentially increasing steps, with the aim to avoid heating problems. CFM current-voltage measurements have been performed by delta mode four probe technique ad-hoc modified in order to minimize possible heating effects. In this configuration, the current bias is pulsed; each pulse is in rectangular shape with a power-on time of 100 ms and an inter-pulse spacing of 2 s. The pulse amplitude is increased linearly. The critical current value was defined by a standard 1 $\mu V\ cm^{-1}$ criterion over the noise amplitude, which is about 100 nV for our experimental setup.

## Results

Table 1 summarizes the irradiation fluences of all the different Hall bars of each sample and the corresponding calculated displacement per atom (d.p.a.) values and average distance between defects. Hall bars, irradiated with different fluences, are indicated by a name composed of a letter, which refers to the sample and a number, which corresponds to the d.p.a. induced by irradiation. It is worth noticing that the same fluence can correspond to different d.p.a. values depending on the energy of protons impinging on the film.

For all samples, there is a pristine reference: for patterned samples there is a pristine Hall bar which is a not irradiated bar while samples used for X-ray diffraction analysis were also measured before irradiation. It is important to have pristine data as a reference of the properties of the films before irradiation. Indeed, for PLD deposited films, the sample-to-sample variability of transport properties may be comparable to the effects of irradiation that we are investigating. In the forthcoming figures, we adopt the following legend: symbol shapes identify the sample, while grayscale is a measure of the irradiation dose, from empty symbols for the unirradiated Hall bars to increasingly dark colour for increasing dose.

| Sample | Hall bar | Fluence ($10^{16}$ cm$^{-2}$) | Displacement per atom (d.p.a) | Inter-defect distance (nm) |
|---|---|---|---|---|
| A | A_0 | 0 | 0 | --- |
|   | A_0.25 | 0.70 | $2.5 \cdot 10^{-4}$ | 4.3 |
|   | A_0.99 | 2.80 | $9.9 \cdot 10^{-4}$ | 2.7 |
| B | B_0 | 0 | 0 | --- |
|   | B_0.69 | 1.95 | $6.9 \cdot 10^{-4}$ | 3.1 |
|   | B_2.27 | 6.40 | $2.27 \cdot 10^{-3}$ | 2.1 |
|   | B_2.59 | 7.30 | $2.59 \cdot 10^{-3}$ | 2.0 |
| C (with 80 μm Al foil) | C_0 | 0 | 0 | --- |
|   | C_2.30 | 2.68 | $2.30 \cdot 10^{-3}$ | 2.1 |
|   | C_4.59 | 5.35 | $4.59 \cdot 10^{-3}$ | 1.6 |
| D_0 | --- | 0 | 0 | --- |
| D_2.30 (with 80 μm Al foil) | --- | 2.68 | $2.30 \cdot 10^{-3}$ | 2.1 |
| D_4.59 (with 80 μm Al foil) | --- | 5.35 | $4.59 \cdot 10^{-3}$ | 1.6 |
| E_0 | --- | 0 | 0 | --- |
| E_1.90 | --- | 5.35 | $1.90 \cdot 10^{-3}$ | 2.2 |

**Table 1**: Summary of the samples with the relative fluences, average displacements per atom (d.p.a) and distance between defects. Samples A, B and E were directly irradiated with 3.5 MeV protons, while samples C and D were irradiated with protons decelerated through the interposition of a 80 μm thick aluminium foil. The inter-defect distance was evaluated by considering just stable Frenkel pairs defects and neglecting possible clustering and annealing effects. Samples D and E used for X-ray diffraction analysis also after irradiation do not have Hall bar patterns.

The microstructural parameters describing the lattice strain can be calculated by Rietveld refinement, obtaining the Williamson-Hall plots drawn in Figure 3; in particular, these plots provide a qualitative information about the evolution of the lattice strain along [00$l$] as a function of the irradiation. As expected, the pristine samples D and E are characterized by about the same amount of lattice strain; remarkably, after irradiation they display opposite behaviours. In fact, sample E exhibits a decrease of the lattice strain, whereas, conversely, samples pertaining to the D series undergo a progressive increase of the strain with irradiation. This behaviour is clearly related to the different irradiation treatments experienced by these samples; in particular, decelerated protons more effectively produce lattice strain within the thin film.

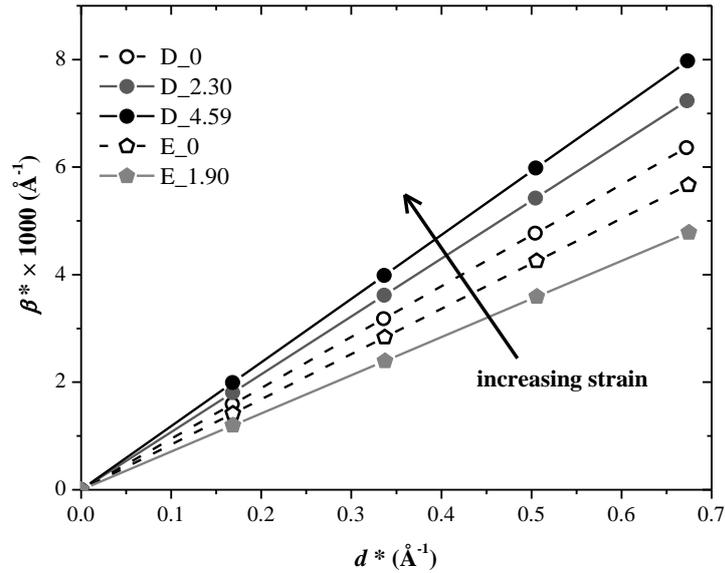

**Figure 3**: Williamson-Hall plot: β (integrated breadth of 100 peak) as function of $d$= (2sinθ/λ), for samples D and E before and after irradiation.

In Figure 4 we show as a reference the critical current density as a function of magnetic field up to 16 T measured on the pristine sample C_0 at 4.2, 8 and 12 K. The $J_c$ values for samples A_0 and B_0 are about 20% lower: at 4.2 K and 7 T $J_c$ is $\approx 2\cdot10^5$ A/cm$^2$ for sample C_0 and $\approx 1.6\cdot10^5$ A/cm$^2$ for samples A_0 and B_0. We outline that we always compared the irradiated samples with its pristine sample, where pristine means a bar measured both before irradiation and after the irradiation process with that bar kept shielded, in order to rule out any spurious or ageing effects.

In Figure 5 and Figure 6 we report the $J_c$ values as a function of the field for the patterns irradiated at the different fluences, where $J_c$ has been normalized to the value of the pristine samples A_0, B_0 or C_0. The magnetic field was applied perpendicular to *ab* crystalline plane and the temperature was fixed at 4.2 K (Figure 5) and at 12 K (Figure 6). Sample A (squares) and sample B (triangles) were measured up to 9 T, while sample C (circles) was measured up to 16 T. As it can be seen samples A and B show an improvement of $J_c$ with the increasing dose. In particular, the bar B_2.27 shows an improvement of $J_c$ of about 40% at 7 T and 4.2 K as compared to its pristine counterpart. The most irradiated bar of sample B (B_2.59) is not reported at 4.2 K because $J_c$ was too high to be measured with our current supplier. At 12 K, as shown in Figure 6, the most irradiated bar of sample B (B_2.59) shows an improvement of $J_c$ of about 50% at 7 T. On the contrary, for sample C, an opposite trend is observed, namely the most irradiated bar C_4.59 shows the worst $J_c$ both in self-field and in-field at all the investigated temperatures. The decrease of $J_c$ at 7 T is of about 80% at 4.2 K and almost 90% at 12 K. Hence, from these plots, $J_c$ does not seem to have a monotonic and unique response to irradiation: samples A and B show an enhancement of $J_c$ with increasing dose of irradiation, whereas sample

C shows the opposite trend. Moreover, we observe that at 4.2 K the in-field behaviour of $J_c$ ratios is flat for all the three samples, while at 12 K $J_c$ ratios increases with increasing field for samples A and B (where we observe an improvement of $J_c$ with irradiation) and decreases with increasing field in samples C (where $J_c$ decreases upon irradiation).

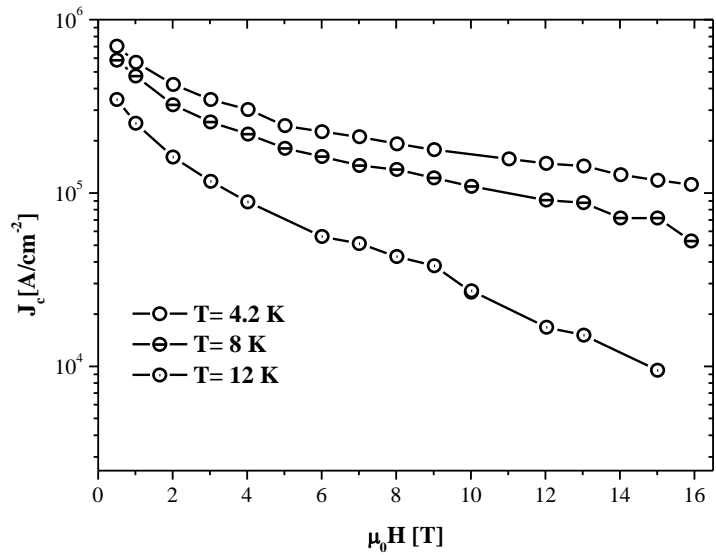

**Figure 4**: $J_c$ vs magnetic field up to 16 T of the pristine sample C at 4.2, 8 and 12 K.

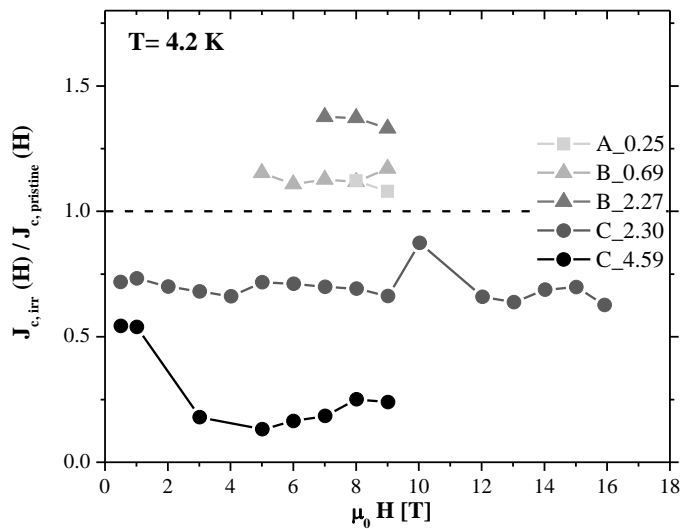

**Figure 5**: $J_c$ vs magnetic field at 4.2 K for the different irradiated patterns normalized to the values of the relative pristine samples at the same fields.

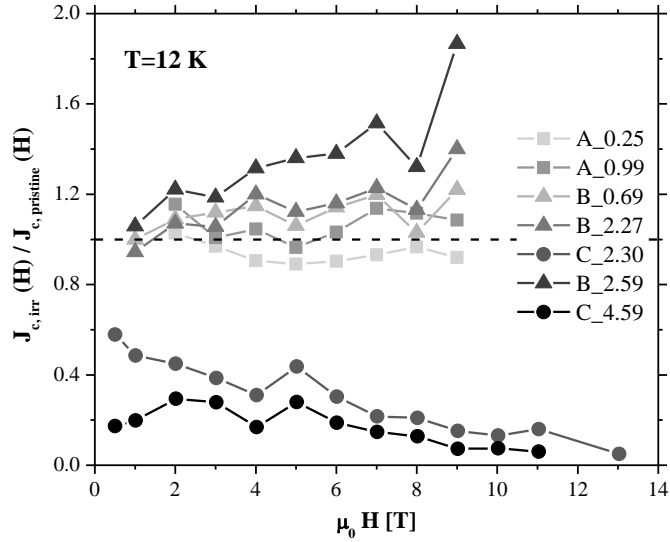

**Figure 6**: $J_c$ vs magnetic field at 12 K for the different irradiated patterns normalized to the values of the relative pristine samples at the same fields.

For a better evaluation of the effects of proton irradiation on $J_c$, we analysed $J_c$ values at fixed temperatures and fields as a function of the d.p.a., in order to take into account not only the number of particles that hit the target but also the expected damage they induced in the film in dependence on their energy. Figure 7 and Figure 8 show $J_c$ values, normalized to the value of the corresponding pristine samples A_0, B_0 or C_0, as function of d.p.a. for all the irradiated bars at 4.2 K and 12 K and at 5 and 9 T. From such plots we observe an increase of $J_c$ for d.p.a values up to about 0.002 relative to the bars irradiated on sample B: at 9 T $J_c$ increases of 33% at 4.2 K (bar B_2.17) and 86% at 12 K (bar B_2.47). On the contrary, sample C shows a decrease in $J_c$ reaching values of about 75% at 4.2 K and above 90% at 12 K and 9 T, when d.p.a. is over 0.004 (bar C_4.78).

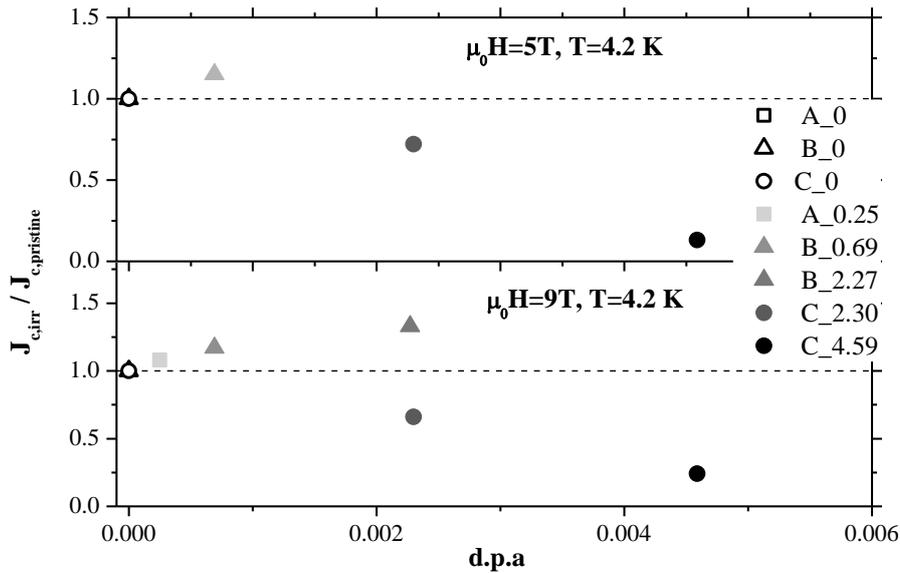

**Figure 7**: $J_c$ of the irradiated bars normalized to the values of the relative pristine samples at 4.2 K and 5 T (upper panel) and 9 T (lower panel).

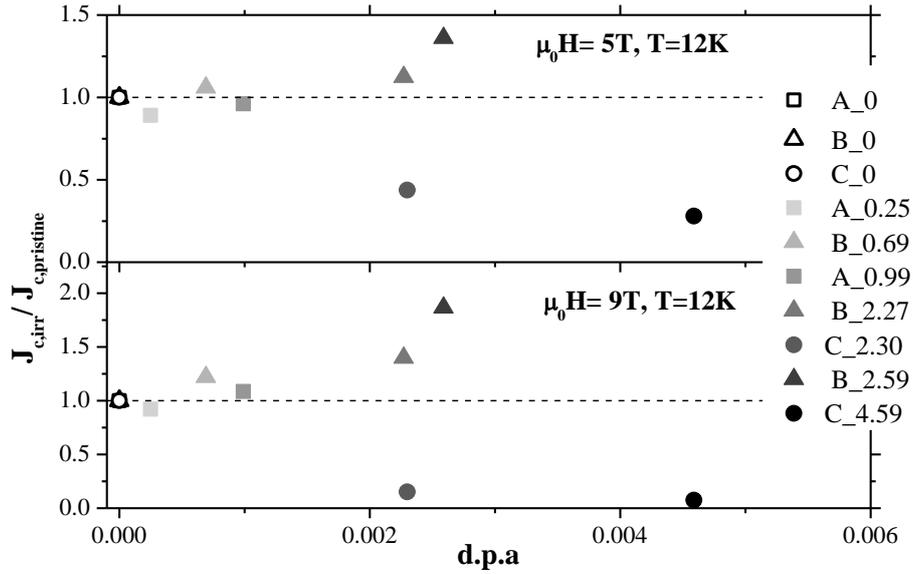

**Figure 8**: $J_c$ of the irradiated bars normalized to the values of the relative pristine samples at 12 K and 5 T (upper panel) and 9 T (lower panel).

In order to evaluate the effect of irradiation on the critical temperature we performed resistivity measurements. In Figure 9 we report the resistive transitions for all the bars of the three samples, where the resistivity is normalized to the value of the respective pristine samples at 20 K. We can see that the variations are very small for samples A and B. Only sample A shows a very little decrease of $\rho(20\ K)$ of about 5% upon irradiation. Sample C, on the contrary, shows a significant increase in the resistivity of about 20%.

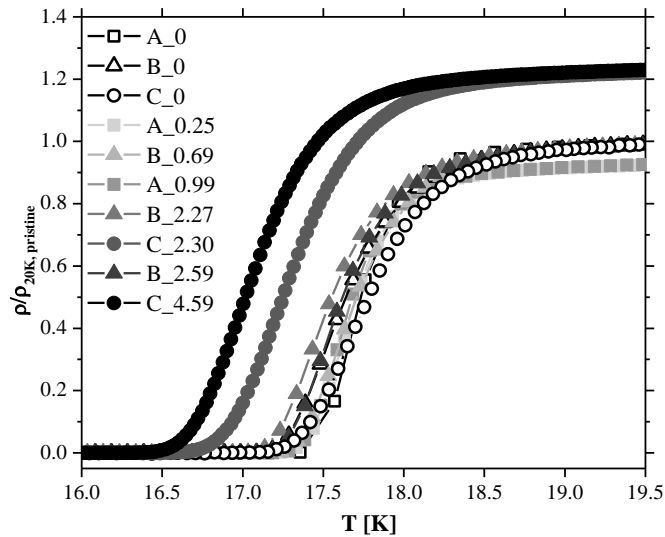

**Figure 9**: Resistive transitions for all irradiated bars of samples A, B and C where the resistivity at 20 K for each sample is normalized to the resistivity of the relative pristine sample at 20 K.

In Figure 10 we report the resistivity in the normal state (20 K), the 90% of the resistive transitions in 0 field and the ΔT$_c$ (T$_c$(90%)-T$_c$(10%)) for all the irradiated bars, all normalised to the pristine values. As already shown from the resistive transitions, we observe that there is no significant variation of the normal state resistivity nor in T$_c$ or in ΔT$_c$ in irradiated samples A and B with respect to the relative pristine sample. In sample C, T$_c$ decreases by about 0.6 K and, at the same time, the resistivity increases by about 20% upon high irradiations.

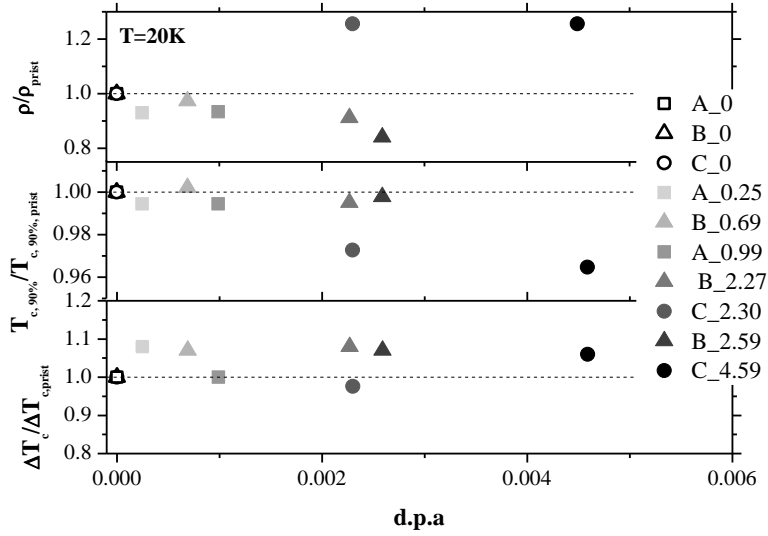

**Figure 10**: Resistivity at 20K (top panel), critical temperature (intermediate panel) and ΔT$_c$ (bottom panel) as function of d.p.a. All the quantities have been normalized to the corresponding value of the pristine sample.

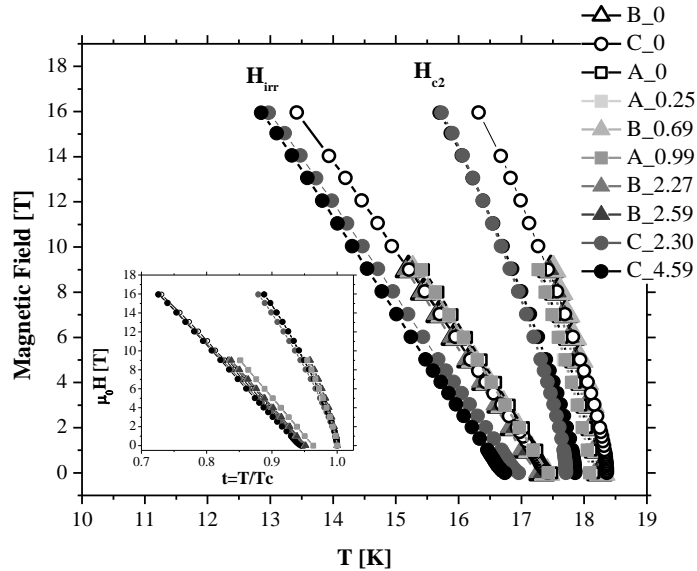

**Figure 11**: Upper Critical Field and Irreversibility field as a function of temperature for all samples. The inset shows upper critical field and irreversibility field of all the bars as a function of the reduced temperature t=T/T$_c$ of each pristine sample.

From resistivity measurements in applied field we evaluated the irreversibility field $H_{irr}$ and upper critical field $H_{c2}$ using the criterion of 10% and 90% of the resistivity value in the normal state above the transition. In Figure 11 we report $H_{c2}$ and $H_{irr}$ for all the bars, evaluated with H // c. In the inset of Figure 11, $H_{c2}$ and $H_{irr}$ as a function of the reduced temperature of each pristine sample are reported. All the curves show very high slopes near $T_c$, as already reported for 11 thin films [23, 16]. Irradiation does not change significantly the slope of $H_{irr}$ and $H_{c2}$ curves: indeed, the curves belonging to the same sample are superimposed over each other for samples A and B, while sample C only shows a slight decrease in $T_c$, as already mentioned above.

**Discussion**

We have analysed five Fe(Se,Te) thin films deposited on $CaF_2$ substrates, which were exposed to proton irradiation at different fluences. For some samples the energy of protons impinging on the film and the implantation depth in the substrate were decreased by the interposition of an Aluminium foil.

As for superconducting properties, the critical fields remain unchanged regardless different fluences and the presence of the Aluminium foil. By contrast, the resistivity and the critical temperature stay substantially unaltered for samples irradiated without Aluminium foil, while the slowing down of protons caused a resistivity increase of about 20% and a suppression in $T_c$ of about 0.6 K. For comparison, a $T_c$ suppression of 2 K has been reported in $Ba(Fe_{0.925}Co_{0.075})_2As_2$ single crystals irradiated with 3 MeV protons and a fluence of $1.2 \cdot 10^{16}$ cm$^{-2}$ [24] and up to 4.3 K in $Ba_xK_{1-x}Fe_2As_2$ single crystals irradiated with 3 MeV protons and a fluence of $9.2 \cdot 10^{16}$ cm$^{-2}$ [25].

Regarding the critical current density, the samples A and B irradiated directly under the 3.5 MeV proton beam, where the protons implant into the substrate several tens microns away from the film-substrate interface (see Figure 2), were studied for d.p.a. up to 0.0026. In these conditions, we observe an enhancement of $J_c$ upon irradiation. For instance, we measure an improvement of $J_c$ of about 40% at 4.2 K and 7 T and up to 50% at 12 K with respect to the pristine bars (see Figure 5 and Figure 6). For sample C irradiated with the interposition of an aluminium foil, which was studied only for d.p.a higher than 0.002, we observe a $J_c$ decrease after irradiation up to 80% at 4.2 K and of 90% at 12 K and 7 T. We point out that in our experiment, in samples irradiated without Al foil, we do not observe $J_c$ to reach a maximum; on the other hand, in samples irradiated with Al foil, $J_c$ monotonically decreases. Hence, the apparent non-monotonic trend of $J_c$ with increasing d.p.a., i.e. first increase and then decrease, appears only by joining data from differently strained samples, also due to the fact that we did not perform experiment in the whole d.p.a. range in a single sample. Therefore, we can speculate that, not only the d.p.a., but also the position of the defects in the substrate can influence the properties of the film and therefore it is able to tune $J_c$. Indeed, the different behaviour observed in samples A and B with respect to samples C cannot be interpreted in terms of d.p.a. alone. It appears that the interposition of a thin Al foil, which reduces the implantation depth of protons into the substrate, actually modifies the film properties via strain in a different way, depending whether the substrate is mainly damaged close to or far from the substrate-film interface. In other words, for equal d.p.a., the closer to the interface are the irradiation defects in the substrate, the larger is the strain in the film and the stronger is the detrimental effect on its superconducting properties. This scenario is supported by the XRD analysis, from which it appears that the shift of the implantation peak close the interface causes an increase in the strain of the films.

Noteworthy, the $J_c$ enhancement upon irradiation is modest as compared to other FeSCs families. In ref. [3] it was pointed out that this can be ascribed to the lower depairing current in 11 FeSCs, which sets the magnitude of achievable currents. The moderate $J_c$ enhancement can be also ascribed to the natural presence of pinning centres in Fe(Se,Te) thin films induced by the growth on the $CaF_2$ substrates [26], where $J_c$ already reaches $10^6$ A/cm$^2$, in self-field at 4.2 K which is about 5% of $J_d$ [3]. The effect of irradiation in the same conditions (3.5

MeV protons), is modest: as a comparison, after irradiation, YBCO thin films show a suppression of $J_c$ of about 20% for a d.p.a. as low as $2.04 \cdot 10^{-4}$ [27]. Generally, the low sensitivity of superconducting and normal state properties of this compound to irradiation may result from the balanced competition of positive and detrimental mechanisms. For example, the distribution of irradiation defects may create a coexistence of areas of tensile and compressive stress, having opposite effects on the superconducting properties, thus resulting in an overall net effect that is small. Only at the largest irradiation doses and in presence of a reduced implantation depth in the substrates that can amplify the irradiation-induced effect the detrimental mechanisms tend to prevail. This robustness of Fe(Se,Te) to irradiation damage is promising for application in magnets for high-energy accelerators.

**Conclusions**

We have conducted a systematic study of the effects of irradiation with 3.5 MeV protons on the superconducting properties (critical temperature, fields and currents) of Fe(Se,Te) thin films grown by PLD on $CaF_2$ substrates. Fluence was varied in a wide range from $0.7 \ 10^{16}$ to $7.3 \ 10^{16}$ cm$^{-2}$; moreover, in order to address the issue of the role of the defected substrate where the protons implant in further modifying the film properties, we irradiated some samples with lower proton energies at comparable d.p.a. values by the interposition of a thin Al foil of a suitable thickness. This allows to control the implantation depth of protons into the substrate, and to check whether the film properties change in a different way when the substrate is mainly damaged close to or far from the substrate-film interface.

It turns out that when protons implant far from the superconducting film, $J_c$ enhancements in FeSe$_{0.5}$Te$_{0.5}$ can be obtained, even if they are not as large as in the case of other FeSCs [3]. In such conditions, critical temperature, critical and irreversibility fields are virtually unchanged, for the employed fluences. Conversely, when the implantation layer is close to the substrate-film interface, both $T_c$ and $J_c$ show a decrease together with an enhancement of the resistivity, pointing to the crucial role of the substrate itself and its possible modifications in determining the superconducting film properties. This should be carefully considered in every irradiation experiment of thin-film superconductors.

Finally, what emerges from our irradiation experiment with high energy protons up to very high values of d.p.a. is that the family Fe(Se,Te) is very robust against proton irradiation with respect to other superconductors, e.g. cuprates, making such superconductor very interesting for applications in harsh environments, where strong radiation emissions are expected, such as in accelerators.